\documentclass[aps,prl,twocolumn]{revtex4}
\usepackage{epsfig}
\usepackage{multirow}
\usepackage{graphicx}
\usepackage{subfigure}
\usepackage{dcolumn}
\usepackage{amsmath}

\begin{document}

 \title {Charge carrier induced lattice strain and stress effects on As activation in Si}

 \author{Chihak Ahn$^1$ and Scott T. Dunham$^{1,2}$}

 \affiliation{$^1$Department of Physics, University of Washington, Seattle, WA 98195 \\
$^2$Department of Electrical Engineering, University of Washington, Seattle, WA 98195}

 \date{\today}

\begin{abstract}

We studied lattice expansion coefficient due to As using density functional theory 
with particular attention to separating the impact of 
electrons and ions.
Based on As deactivation mechanism under equilibrium conditions,
the effect of stress on As activation is predicted.
We find that biaxial stress results in minimal impact on As activation, which is consistent with 
experimental observations by Sugii {\em et al.} [J. Appl. Phys. {\bf 96}, 261 (2004)] and Bennett {\sl et al.}
[J. Vac. Sci. Tech. B {\bf 26}, 391 (2008)].
\end{abstract}

\maketitle

Stress  effects  are of great interest in  modern  ULSI  technology  since  they  can  be 
employed  to  improve  various  material  properties.  Uniaxial  stress  has  been  employed  in 
MOSFET  devices  since  the  90  nm  node  technology  step  to  improve  carrier  mobility~\cite{Thompson}.  
Properly  applied  stress  can  also  suppress  dopant  diffusion~\cite{Moriya, Fang, Ahnmrs06},  
enhance  activation~\cite{Fang, Lever, AhnJVST},  and reduce the band gap~\cite{Yang}. 
Therefore, understanding stress effects becomes essential for further MOSFET scaling. 

As deactivation is governed by As$_m$V$_n$ cluster formation, and clusters with $m=1-4$ and $n=1$ are considered 
as the dominant species in deactivation kinetics~\cite{Berding}. 
Under equilibrium conditions, the concentrations of defect X (As, V or As$_m$V)
is determined by the free As and V concentrations and cluster formation energies: 
$C_X=A\exp\left(-E_X^f/kT\right)$,
where $E_X^f$ is the formation enthalpy, and A includes the configuration and formation entropy.
The total chemical As concentration is given by
\begin{eqnarray}
C^{total}_{As}=C_{As}+\sum_{m=1}^4 m C_{As_mV}.
\label{As:Ctotal}
\end{eqnarray}
Table~\ref{As:EF} lists the formation energies of As$_m$V complexes based on
the total free energy of 64-atom (or 63-atom, with vacancy) super-cells using the density functional theory 
(DFT) code VASP~\cite{Kresse} with PW91 GGA potential~\cite{PW91}.  
All calculations were done at a 250 eV energy cut-off with 2$^3$ 
Monkhorst-Pack {\bf k}-point sampling~\cite{Monk}. 
Each time an As atom is added to a vacancy, the formation energy 
is lowered by about $1.5$ eV, and thus a larger complex is more stable than a smaller one.
We calculated the As$_m$V concentrations based on the formation energies listed in Table~\ref{As:EF}.
Since DFT GGA underestimates the vacancy formation energy by about 1 eV~\cite{Batista}, 
we also applied a correction for the As$_m$V formation energies using experimental values~\cite{UralPRL}.   
As$_4$V has the lowest formation energy and becomes the dominant cluster under equilibrium conditions. 
Smaller clusters can be 
formed during epitaxial As-doped Si growth and early stages of annealing, and can dominate before full 
equilibration is reached~\cite{Ranki,Derdour}, but we restrict our analysis to equilibrium conditions. 

\begin{figure}[tbhp]
 \begin{center}
 \includegraphics[trim = 0mm 0mm 0mm 1mm, clip, width=7.0cm, angle=0]{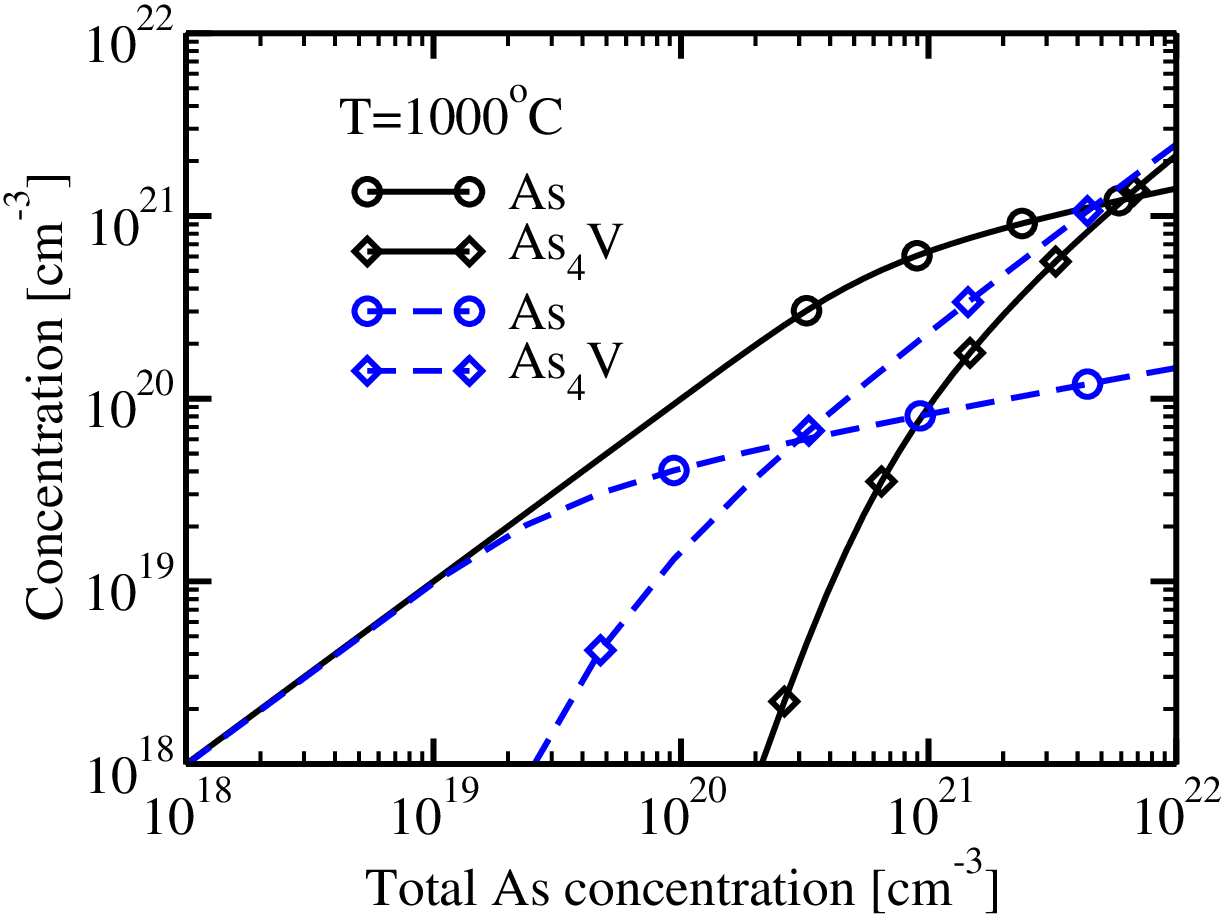}
  \end{center}
  \vspace{-0.13in}
  \caption{Equilibrium As concentration and As$_4$V concentration as a function of the total chemical As concentration for $E_F=E_c$. 
   Solid lines are plotted with correction for vacancy formation energy and broken lines are plotted with DFT 
   formation energies. Smaller clusters  don't appear due to low concentration.}
  \label{As:CAs}
\end{figure}

Fig.~\ref{As:CAs} shows the isolated As concentration as a function of the total As concentration. As the number of As forming 
As$_4$V increases to become a significant fraction of free As, the free As concentration starts deviating from the total chemical 
As concentration, which is consistent with previous reports~\cite{Erbil, Derdour}. We should note that the As$_4$V formation energy is 
actually Fermi level dependent due to a charge transfer from the Fermi level to the cluster when As$_4$V forms. 
A higher Fermi level 
results in lower cluster formation energies, and thus the As$_4$V (As) curve becomes steeper (flatter) when the Fermi level 
dependent formation energy is used. Fig.~\ref{As:CAs} uses $E_F=E_c$, which is appropriate for degenerately doped Si.
\begin{table}[bt]
  \vspace{-0.1in}
  \caption{Formation energy of As$_m$V clusters. When the experimental vacancy formation energy is used (4.60 eV~\cite{UralPRL}),
  formation energies increase by about 1 eV. The experimental value of the V formation energy was calculated by 
  subtracting the migration barrier (0.26 eV, DFT value) from the activation enthalpy (4.86 eV~\cite{UralPRL}).
  In the second row, the first value is based on the DFT result, and the second is based on the experimental V formation energy. }
 \begin{center}
  \begin{tabular}{cccccc}
  \hline
        & V& AsV  & As$_2$V & As$_3$V & As$_4$V \\
  \hline
   \multirow{2}{1in}{E$^f$ (eV)} &3.59 & 2.15 & 0.68 & -0.66 & -2.22 \\
               &4.60 & 3.16 &1.69 &0.35 &-1.21\\
  \hline
  \end{tabular}
  \label{As:EF}
 \end{center}
\end{table}

In equilibrium, the change in the concentration of a defect X due to stress is given by 
\begin{eqnarray}
\frac{C_X(\vec\epsilon)}{C_X(0)}\approx\exp\left(-\frac{\Delta E^f_X(\vec \epsilon)}{kT}\right),
\label{As:Cepsilon}
\end{eqnarray}
where $\Delta E^f_X(\vec \epsilon)$ is the change in the formation energy of X due to stress.
In the case of the As$_m$V cluster, it is given by~\cite{Diebel} 
\begin{eqnarray}
\Delta E^f_X(\vec \epsilon)=-V_0(\Delta\vec\epsilon_{As_mV}-m\Delta\vec\epsilon_{As})\cdot{\bf C}\cdot\vec\epsilon,
\label{As:DEF}
\end{eqnarray}
where $V_0$ is the volume of a lattice, $\Delta \vec\epsilon_{As_mV}$ ($\Delta\vec\epsilon_{As}$) is the induced strain 
due to As$_m$V (As), {\bf C} is the elastic stiffness tensor of Si, and $\vec\epsilon$ is applied strain.
The induced strain can be determined from the energy vs. strain curve.
A detailed explanation can be found in Ref.~\cite{Diebel}. 
The results are summarized in Table~\ref{As:IS}.
We repeated calculations using 216 supercell for selected structures and acquired the same induced strains.

\begin{table}[tb]
\vspace{-0.13in}
  \caption{Induced strain for As and As$_m$V complexes. As produces small lattice expansion and 
  As$_m$V complexes result in lattice contraction.}
 \begin{center}
  \begin{tabular}{ccccccc}
  \hline
        & ~ As&V &~AsV  &~ As$_2$V &~ As$_3$V & ~As$_4$V \\
  \hline
   $\Delta\epsilon$&~ 0.018&~ -0.25 &~-0.21 &~ -0.22 &~ -0.11 &~ -0.08 \\ 
  \hline
  \end{tabular}
  \label{As:IS}
 \end{center}
\end{table}
\begin{table}[tb]
\vspace{-0.13in}
  \caption{Induced strain due to As, As$^+$, and free electrons and holes. The numbers in parenthesis are extracted from 
  Cargill {\em et al.}~\cite{Cargill}. Note that in spite of longer As-Si bond length
  in Si$_{63}$As$^+$ supercell (Table~\ref{As:AsBond}), the lattice undergoes contraction.}
 \begin{center}
  \begin{tabular}{ccccc}
  \hline
        & ~ As$^0$ &~As$^+$  &~ e$^-$ &~h$^+$\\
  \hline
   $\Delta\epsilon$& ~0.018 (-0.019)&~ -0.22 (0.07) & ~0.22 (-0.09) &~ -0.26 \\
  \hline
  \end{tabular}
  \label{As:IScharge}
 \end{center}
\end{table}
\begin{table}[tb]
\vspace{-0.13in}
  \caption{Local lattice structure around an As atom in the Si lattice compared to atomic spacing in pure Si.}
 \begin{center}
  \begin{tabular}{ccccc}
  \hline
        &~~  Si & ~~As$^0$ &~~ As$^+$ &~~ As (exp)\cite{Koteski} \\
  \hline
  1NN &~~ 2.36 & ~~2.45 & ~~2.43 &~~ 2.43 \\
  %\hline
  2NN &~~ 3.86 &~~ 3.87 & ~~3.86 &~~ 3.87 \\
  %\hline
  3NN & ~~4.53 &~~ 4.53 &~~ 4.52 &~~ 4.53 \\
  \hline
  \end{tabular}
  \label{As:AsBond}
 \end{center}
\end{table}

\begin{figure}[tb]
 \begin{center}
 \includegraphics[trim = 0mm 0mm 0mm 1mm, clip, width=7.0cm, angle=0]{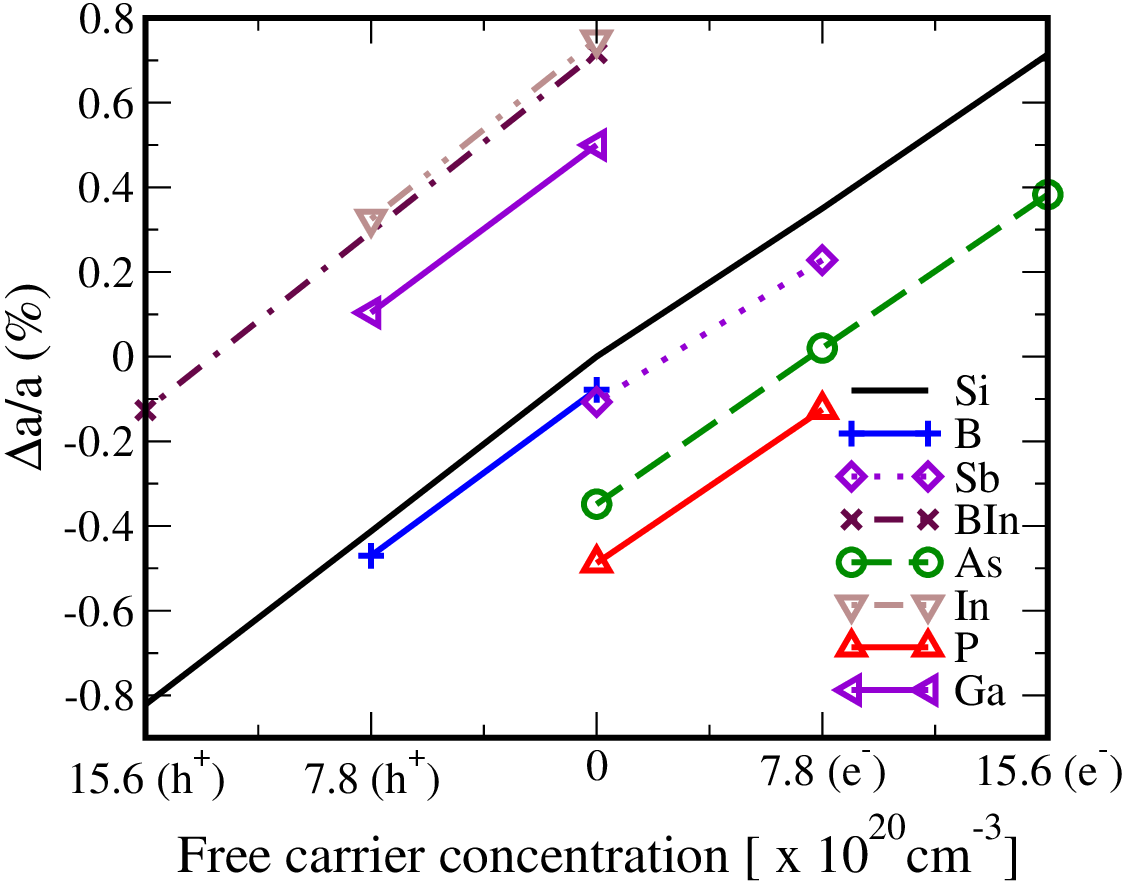}
  \end{center}
  \vspace{-0.13in}
  \caption{Change in lattice constant due to free charge carriers. The lattice undergoes expansion 
   (contraction) as free electrons (holes) are added. Induced strains are obtained by finding equilibrium 
    lattice constant of charged supercell with various dopants. One electron in a 64-atom supercell corresponds 
   to $7.8\times 10^{20} {\rm cm^{-3}}$.}
  \label{As:straincharge}
\end{figure}

As shown in Table~\ref{As:IS}, As is calculated to give a small lattice expansion.
However, several authors have observed lattice contractions in heavily As-doped Si, which they attributed to 
free electrons in the conduction band~\cite{Cargill, Parisini, Herrera}.
In contrast to their conclusion, DFT calculations predict a lattice expansion due to free electrons in the conduction band (Table~\ref{As:IScharge}).
In Cargill {\em et al.}, the total induced strain ($\Delta\epsilon_{As}=\beta_{\rm total} N_{As}$) is 
assumed to be given by the sum of the induced strain due to 
ions ($\Delta\epsilon_{As^+}=\beta_{\rm size}N_{As}$) and free electrons ($\Delta\epsilon_e=\beta_e N_{As}$). 
As shown in Table~\ref{As:IScharge}, the calculated induced strain due to As$^{\rm 0}$ has opposite sign to
measured value, but 
the absolute difference is small and thus its impact on stress effects is  minimal.
However, the reasoning is very different in each case, which raises a fundamental question about the role of electrons: 
Do electrons cause expansion or contraction in the lattice?
To answer this question, we performed DFT calculations to find equilibrium lattice constants of
charged supercells. From the carrier concentration vs. change in lattice constant 
(Fig.~\ref{As:straincharge}), we conclude that electrons expand the lattice while holes cause lattice contraction.

\begin{figure}[tb]
 \begin{center}
 \includegraphics[trim = 0mm 0mm 0mm 1mm, clip, width=7.0cm, angle=0]{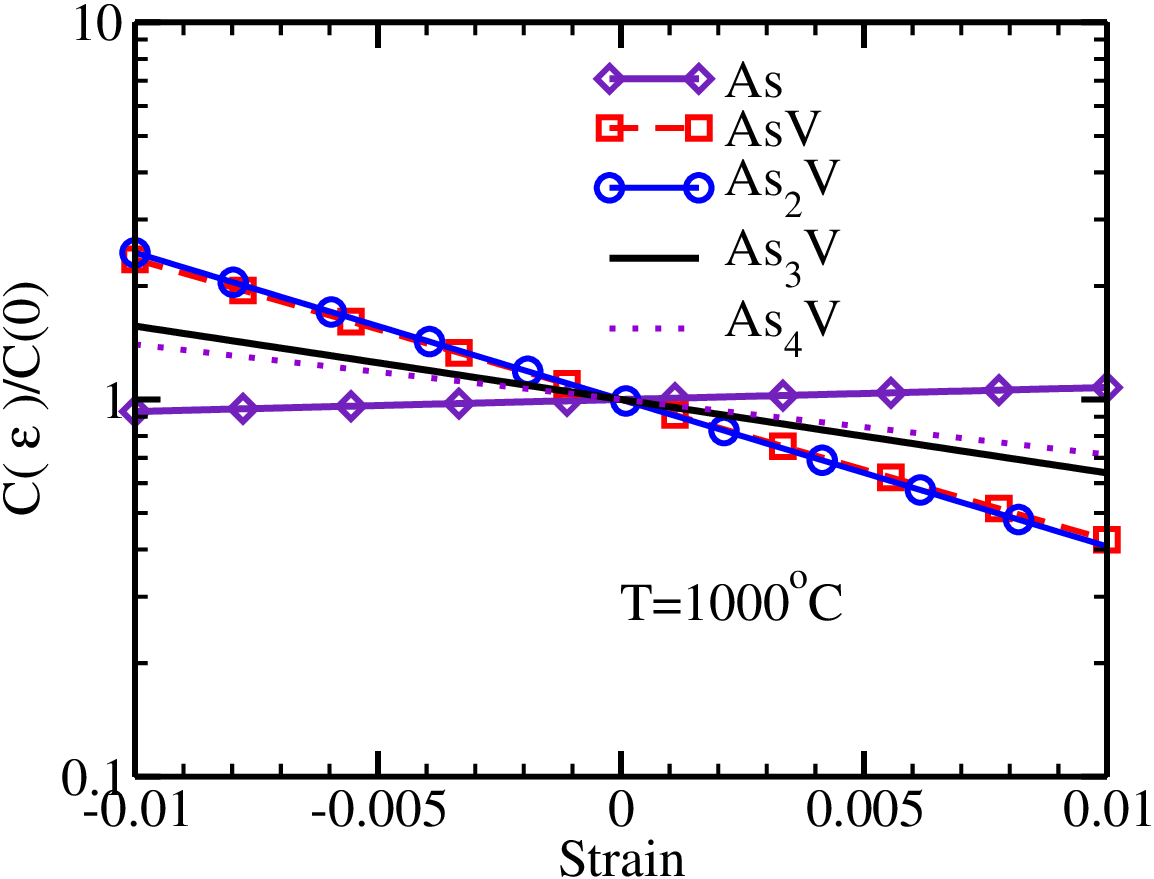}
  \end{center}
  \vspace{-0.13in}
  \caption{Stress effects on As and As$_m$V cluster concentration under biaxial stress. Note that the two dominant 
   complexes, As and As$_4$V, have minimal stress effects.}
  \label{As:Cstress}
\end{figure}

The lattice expansion due to electrons raises another question about the relation between Si-As bond length 
and the lattice parameter. We looked into the local structure around As in Si matrix to answer this question.
As listed in Table~\ref{As:AsBond}, DFT calculations agree with experimental measurement 
up to the 3NN distance and predict a local volume expansion around As~\cite{Koteski, Wei, Erbil}. However, this expansion is attenuated 
as distance increases and As-Si 3NN spacing is very similar to Si-Si 3NN distance. Therefore, changes in the 1NN bond 
length are not directly linked to changes in the lattice parameter, and care should be taken when linking short range atomic 
spacing to lattice constant. In fact, As$^+$ produces a lattice contraction ($\Delta\epsilon=-0.22$) in spite of longer As-Si bond length.
A free electron in the conduction/impurity band overcompensates this contraction, and thus neutral As results in an overall tiny 
expansion ($\Delta\epsilon=0.018$). 

Based on the our analysis, it is likely that experimentally observed lattice contractions originate from reasons other than free electrons. 
We attribute them
to high concentrations of vacancies in the form of As$_m$V$_n$ clusters, and find that a vacancy concentration of about 15\% of the 
As concentration can reproduce 
the lattice contraction observed by Cargill et al.~\cite{Cargill}. 
%Even lower vacancy concentrations (8\%) relative to As give the same effects when 3$^3$ {\bf k}-point sampling is used. 
This level of vacancy concentration was reported based on {\em ab-initio}
calculations by Berding {\em et al.}~\cite{Berding} and positron annihilation spectroscopy by Borot {\em et al.}~\cite{Borot}.
\begin{figure}[tb]
 \begin{center}
 \includegraphics[trim = 0mm 0mm 0mm 1mm, clip, width=7.0cm, angle=0]{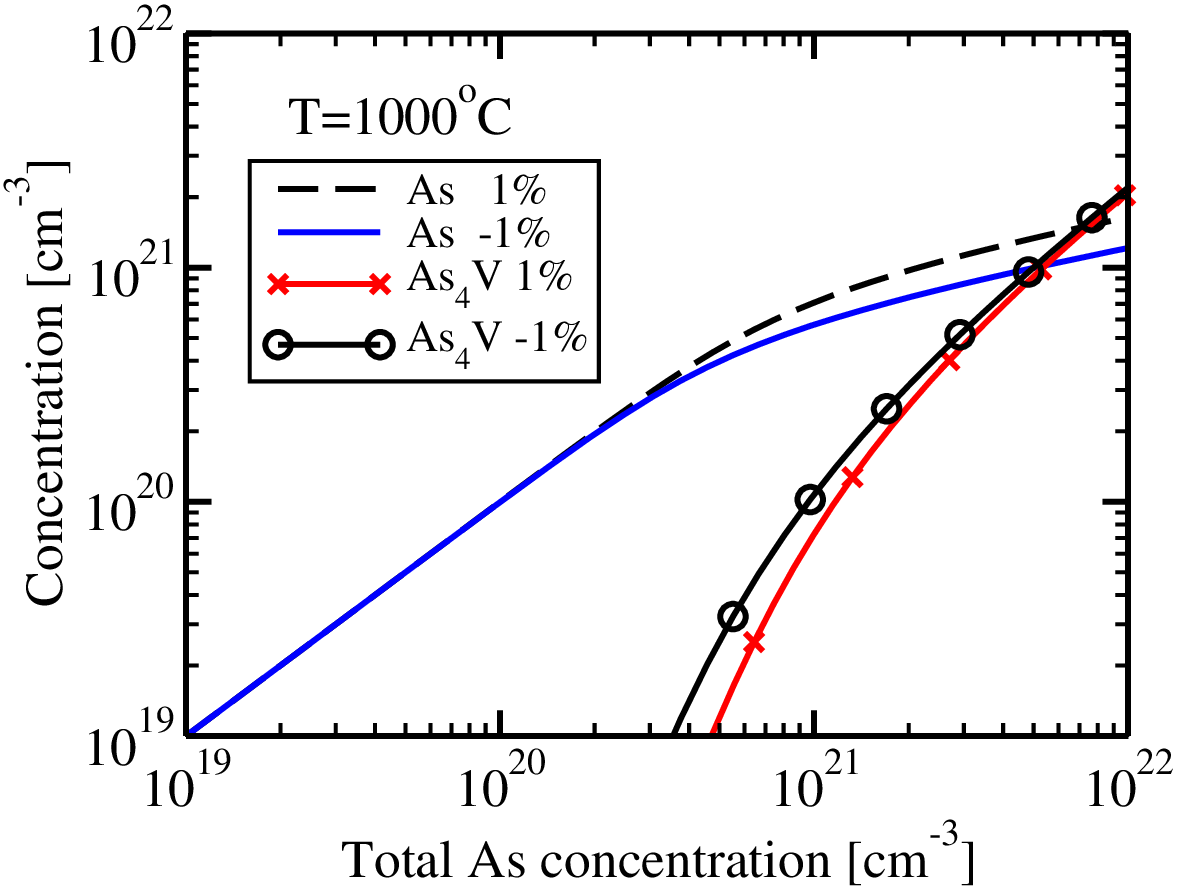}
  \end{center}
  \vspace{-0.13in}
  \caption{Stress effects on As and As$_m$V cluster concentration under biaxial stress. Note that two dominant 
   complexes, As and As$_4$V, have minimal stress effects.}
  \label{As:Asstress}
\end{figure}

Effects of stress on As and As$_m$V concentrations are plotted in Fig.~\ref{As:Cstress} based on Eqs.~\ref{As:Cepsilon} and \ref{As:DEF}. The 
concentrations of the two dominant configurations, As and As$_4$V, undergo changes in opposite directions under biaxial stress, 
but the magnitude is minimal due to the small induced strain. Finally, the free As concentration as a function of 
the total As concentration is plotted in Fig.~\ref{As:Asstress}. At a given total As concentration, compressive biaxial stress 
enhances As$_m$V formation, and thus the number of active As decreases. However, stress effects are minimal due to 
the small induced strains of dominant structures, in accordance with previous experiments~\cite{Sugii,Bennett}.

In conclusion, by performing DFT calculations of the local structure around As in the silicon lattice,
we found that lattice expansion due to the larger size of an As atom is limited to within 3NN distances.
The lattice contraction in highly As-doped Si can be explained by 
As$_m$V cluster formation rather than free electron as previously suggested~\cite{Cargill}. 
The small induced strain due to both isolated As and the dominant deactivated cluster As$_4$V results in negligible stress effects 
on the carrier concentration, in accordance with experimental observations~\cite{Sugii,Bennett}

\section{Acknowledgments}
This research was supported by the Semiconductor Research Corporation (SRC). 
Computer hardware used in this work was provided by donations from Intel 
and AMD as well as NSF award EIA-0101254.

{}
\end{document}